# A Survey on backup of data on Remote Server

**Manali Raje[1], Debajyoti Mukhopadhyay[2]**
[1]Department of Information Technology, Maharashtra Institute of Technology
Pune, India
*manali04raje@gmail.com*

[2]Department of Information Technology, Maharashtra Institute of Technology
Pune, India
*debajyoti.mukhopadhyay@gmail.com*

**Abstract**: Large amount of electronic data is generated in Cloud computing every day. Efficient maintenance of this data requires proper services. Hence a method to collect data securely, by protecting and developing backups is mentioned. The Objective is to provide Auto Response Server, better solutions for data backup and restoring using Cloud. Data can be collected and sent to a centralized repository in a platform independent format without any network consideration. This data can then be used according to the requirement. The purpose of this particular Remote Backup Server is to collect information from any remote location even if network connectivity is not available at that point of time and provide proper services as well as to recover data in case of loss.

## 1. Introduction

Cloud Computing has evolved over time and has become the most widely used Technology. Cloud has moved ahead of all the previous technologies of Computing be it the Services provided, the methods used for storing data etc. Its advantages have overcome the disadvantages of the techniques used before and the need and use of Cloud Computing is increasing day-by-day. It offers quick services with minimum efforts. The Best Feature offered by Cloud Computing is the Online Data Storage. The data stored on cloud is very sensitive. It belongs to different fields of Medical Science and social networks. There is a need for the user to authenticate himself before he stores the data on the cloud. Also the data stored should not be ill handled The Data is stored in a virtualized pool hosted by the third party. Large Data is operated on by the hosting company using large data centres. The Data from the Virtual Pool is then passed on to the users according to their requirements wherein it can be used for storage of files or data objects.

The Online data storage is used by many users simultaneously and hence there is a possibility that the data can be accessed by anybody. The data stored on the cloud comes at a risk with an occurrence of Human mistake, equipments fault, network connectivity error or any third party's spoofing intentions. The cloud is vulnerable to the Byzantine attacks and this may cause failure in the storage systems. Data modification is also an issue as it is prone to different kind of attacks while changes are being made. Different ways have been proposed to provide authentication on the data. Encrypting the data with the public key is one of the ways. Trust is a major factor which needs to be considered and the cloud or the user should not deny of the performed operations. However the Cloud is changing frequently, which may cause alteration of data stored on Cloud. The Changing of Cloud is known as Data Dynamics. Data Integrity is needed as there are limitations on storage and data back-up. This is due to the fact that the Cloud Server handles large data which does not change during the Storage or transmission

The use of this technology and Computers has become an integrated part of Human Lives. Computers, laptops and tablets are used for storing important data files and other information. Now, if in case the data files get corrupted or if the data is leaked in any circumstances, then to recover the data is an impossible task. In manually maintained systems, the admin has to be contacted where he will search all the records and then present the data back as requested. But with the use of Computers, until and unless there is a backup system, data cannot be retrieved back. Hence a backup facility is needed to restore the data lost from the main server. A solution to the above mentioned problem can be collecting the data and sending it to a centralized storage location in a platform independent way without any network consideration. A Central Storage is used for this purpose where all the client applications will be stored and can be carried to any machine. The stored data should be platform independent. Encryption can be done on the data so that protection can be provided to the data in case of any thefts.

## 2. Related Work

This section puts a light on all the data back-up and recovery techniques developed and used in Cloud Computing. A detailed study shows that, the techniques used before did not provide good performance with respect to cost, security and proper and recovery. A brief comparison of the techniques is given. The PCS Technique is reliable, simple, easy and more efficient for recovery of data. It is based on parity recovery service. Using this method, data can be retrieved with a very high probability. The Parity information is created using the Exclusive-OR, though it cannot control the complexities generated while implementation [2].
 The laptop, smart phone users are handled by the HSDRT technique. It uses ultra-widely distributed data transfer mechanism along with high speed encryption. But the implementation cost is high and also redundancy cannot be controlled [1].
The Efficient Routing Grounded on Taxonomy (ERGOT) Technique [3] is based on semantic analysis but has no focus on time and complexity of implementation. This approach helps for Discovery of Service in cloud computing. ERGOT provides an efficient way based on similarity of semantics to retrieve data. Another method known as the Linux Box method contains less steps for data recovery and backup. It makes easy, the migration of services from one cloud to another.
In Cold and Hot back up strategies, the cost of implementation increases as the data increases [4]. This approach performs recovery and backup based on the basis of failure detection. The services are triggered upon the detection of service failure in Cold Backup approach. The services won't be triggered when available. In Hot Backup approach the services are in the activated state which is a transcendental strategy for service composition in dynamic network.

The authentication works have also been mentioned. The Byzantine attacks may cause the storage system to fail in any way [5].The concept of searchable encryption is discussed wherein the cloud does not know the query but returns what is asked for [7]. The use of cryptographic techniques using public keys is mentioned in [6].

Though these techniques offer solutions; they are not efficient enough to retrieve the data properly. Hence the Remote data back-up server's role is important and thus a topic of discussion along with authentication of the data stored on cloud.

## 3. The Backup Server

These days, "Cloud Sever" seems to be making headlines with anything related to Information Technology. It makes large amount of data and computational resources available through a variety of interfaces. Whenever the data in the laptops, smart phones or any internal system is taken out and put on the data centre which is owned by another company, then they are said to be "moved to the cloud". Client and service providers are the two components of Cloud Server. Whenever we say "a backup" of anything we mean to have a copy or a replacement for it. A Remote Data Backup Server is something which is located far from where the actual data is stored and has the same state as that of the main cloud. The main Cloud can be termed as the Central Storage whereas the Remote backup is known as Remote Storage.

Consider a scenario where the data stored on the cloud is lost due to some reasons say human error or natural calamity. The said approach will be helping to retrieve the data back by fetching the data from the remote storage. This system can be useful to retrieve the data in case of any permanent loss. Data can be retrieved even in case of no network connectivity. The approach is more secure than the other systems, very flexible and easy to use for any user. Also the implementation cost is less as compared to the other systems.

Even if this approach is easy to deal with, there are some issues which need to be considered, such as:

a. Integrity of Data:
    This deals with the State and structure of the data. It checks whether the data remains unchanged during the operations.

b. Security:
    Giving security to the data stored on the cloud is the first concern of the backup server. It should not allow any unauthenticated user to get a glimpse of the data.

c. Confidentiality: The data should be kept confidential as there are many users accessing the cloud. Provision to hide the personalized data should be provided.

d. Availability:
    The cloud services should be available whenever required by the users. Care should be taken that the denial of services is not encountered.

e. Trustworthy:
    The Cloud servers and the users should not in any case deny the operations performed.

f. Consistency:
    The data stored should be consistent and the format should be in accordance to what is required by the storage system.

g. Redundancy:
    Redundancy is a major issue and proper mechanism should be followed to avoid it.

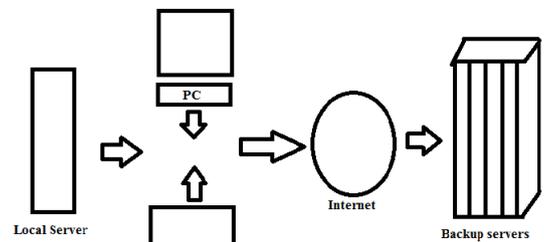

Figure1. Architecture of remote data backup server

### I. PROPOSED PLAN

The Seed Block Algorithm gives an efficient solution to deal with security, implementation cost, complexity of implementation etc. The architecture is as follows:

### 3.1 Architecture

The algorithm used is known as the Seed Block Algorithm. It uses the Exclusive-OR Operation. Suppose there are two files A and B. When A and B are EXORed, the result will be produced and stored in C. Simple backup and recovery is provided. A main cloud server, its clients and the remote data server form the architecture (Figure 1). A random number and a unique client id are set for every client who gets EXORed when registration is made in the cloud. The client id and the random number get EXORed with each other to produce a block for a particular client. This block is known as the seed block, which is stored in the remote server. A file which is created for the first time is stored on the main cloud. While it gets stored in the main server, it is EXORed with the seed block of the related client, which gets stored in the remote server in the form of a file'. Now, if in any case the main cloud crashes or damages or any file is accidentally or purposely deleted, then the original file can be retrieved by EXORing the file stored on the remote server.

### 3.2. Algorithm

The algorithm is as follows:

Initialization: Main cloud: $M\_C$; Remote Server: $R\_S$; Clients of Main Cloud: Cl;

Files: f1 and f1'; Seed block: S; Random Number: r; Client ID: $Cl\_id$

The file f1 created by Cl and the random number r generated at $M\_C$ are the inputs. The output is the recovered file f1 after its deletion from the $M\_C$.

Step 1: int r=random (); //*Generate a random number.*

Step 2: Create a seed block S for each Client on Main Cloud and store it at the remote server.

S= r EXOR $Cl\_id$.

Step 3: If modification is made to the file stored on $M\_C$ then, f1' is created as f1'=f1 EXOR S

Step 4: f1' is stored at the Remote Server

Step 5: If the server crashes it means that f1 is deleted from $M\_C$. To retrieve it from the remote server f1=f1' EXOR S is used.

Step 6: Return f1 to Cl.

Step 7: End

## 3.3 Flowchart

The following figure shows the workflow of the system. The main components are the repository, web service, the database and the users (Figure 2). The application is maintained at the client's side on the laptop or a smart phone. It can be ported to any machine. Data from these devices which is independent of the platform used is sent to the Central repository. Validation of the incoming data is checked and if authenticated is passed on to the virtual database. The database is connected to the users through web services and then the customized data is sent over to the user.

The components of the system perform the following functions:

- Verify the user and the information before storing it onto the database.
- The user's information, the time of updation are also recorded.
- The repository can handle multiple requests at a time.
- The data stored can be shared with other users.

The data is encrypted before storing so that in case of any mishandling the data remains safe and unaltered. Data validation is done before storing it in the database. This approach allows the user to remain confident that the data remains safe in the remote back-up server and can be retrieved in any situation of loss. Security is assured as the stored data is encrypted using encryption techniques. Authentication of the data becomes an important aspect. If the data is not authenticated or the user of the data is not authenticated it can create great risks for the cloud and the other users. The unauthenticated data may contain spam and irrelevant data objects which can prove harmful for the system and the other users leading to ill minded activities.

## 3.4 Users and their Roles

Figure 3 shows the users of the system and the role they play:
The users are classified into two types:
a. Internal users: They may be the administrators or the experts related to a particular domain
b. External Users: They may comprise of any person who is in need of any information, help or requirement gathering.

The data from both types of users is stored in the database. The Admin's job is to maintain the records and data as the requirement demands. It is the job of the admin to take care of the Cloud server and the web services. The restoration can be done either manually or automatically.

Multiple requests from different users can be handled simultaneously through the data processing web server.

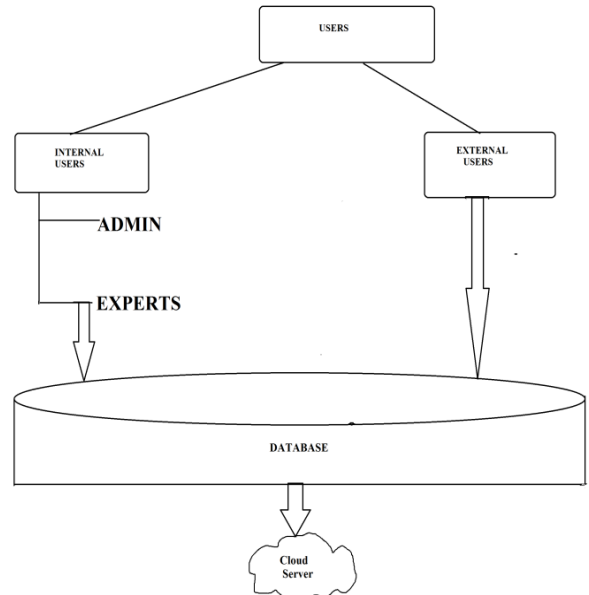

Figure3. Users and their roles

Since it is platform independent data from various platforms is collected and loaded onto the database. Before loading the data verification and validation is carried out so that no harm is meant to the data which is being stored. The Data backup server will present the recovered data if in case the main cloud storage crashes or the data is lost in any case.

The Advantages of using this system is its flexibility, availability, portability, easy maintenance, robustness, easy to use and proper backup facilities etc.
*Flexibility:* The system is flexible enough to add on any new feature. It works efficiently and supports the features added.
*Portability:* Works in any environment and is platform independent.
*Fast*: The System is faster than manually maintained systems and has a low to nil scope for human error made.
*User Friendly*: Presents easy ways to manage the system and also has low cost implementation.
*Reliability*: Is reliable and trustworthy because authentication is provided and security is maintained.
*Proper backup facility*: Provides the same sized data in case of loss from main cloud.

## 4. Conclusion

Due to the ever increasing use of technology, large amount of data is produced and stored. This data is stored securely on the remote servers using the seed block algorithm. The algorithm is also used to retrieve the data if lost in any circumstances. Also we have seen that before storing the data, it is authenticated and validated which prevents unauthorized activities and hence security is maintained. Hence an efficient way to store and retrieve data in a safe manner has been discussed.

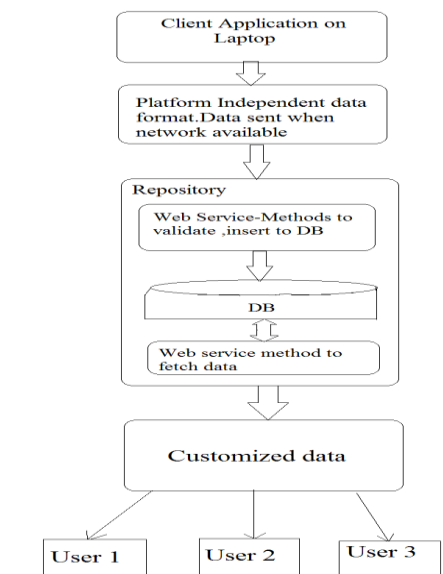

Figure 2: Workflow of the system

## References

[1] Yoichiro Ueno, Noriharu Miyaho, Shuichi Suzuki, Muzai Gakuendai, Inzai-shi, Chiba, Kazuo Ichihara, 2010, "Performance Evaluation of a Disaster Recovery System and Practical Network System Applications," Fifth International Conference on Systems and Networks Communications, pp 256-259.

[2] Chi-won Song, Sungmin Park, Dong-wook Kim, Sooyong Kang, 2011, "Parity Cloud Service: A Privacy-Protected Personal Data Recovery Service," International Joint Conference of IEEE TrustCom-11/IEEE ICESS-11/FCST-11.

[3] Carole Goble, 2010, "ERGOT: A Semantic-based System for Service Discovery in Distributed Infrastructures," 10th IEEE/ACM International Conference on Cluster, Cloud and Grid Computing.

[4] Lili Sun, Jianwei An, Yang Yang, Ming Zeng, 2011, "Recovery Strategies for Service Composition in Dynamic Network, "International Conference on Cloud and Service Computing.

[5] C. Wang, Q. Wang, K. Ren, N. Cao and W. Lou, "Toward Secure and Dependable Storage Services in Cloud Computing", *IEEE T. Services Computing*, vol. 5, no. 2, pp. 220–232, 2012.

[6] H. Li, Y. Dai, L. Tian, and H. Yang, "Identity-based authentication for cloud computing," in *CloudCom*, ser. Lecture Notes in Computer Science, vol.5931. Springer, pp. 157–166, 2009.

[7] S. Kamara and K. Lauter, "Cryptographic cloud storage," in *Financial Cryptography Workshops*, ser. Lecture Notes in Computer Science, vol.6054. Springer, pp. 136–149, 2010